\documentclass[preprintnumbers,amsmath,amssymb,twocolumn]{revtex4}
\usepackage{graphicx}
\usepackage{dcolumn}
\usepackage{bm}
\usepackage{braket}
 \usepackage{mathrsfs}
 \usepackage{physics}
 
 \usepackage{amsmath}

\begin{document}

\title{Diamond nano-pyramids with narrow linewidth \\SiV centers for quantum technologies}
\author{L. Nicolas$^{1}$}
\author{T. Delord$^{1}$}
\author{P. Huillery$^{1}$}
\author{E. Neu$^{2}$}
\author{G. H\'etet$^{1}$} 
\affiliation{$^1$Laboratoire Pierre Aigrain, Ecole normale sup\'erieure, PSL Research University, CNRS, Universit\'e Pierre et Marie Curie, Sorbonne Universit\'es, Universit\'e Paris Diderot, Sorbonne Paris-Cit\'e, 24 rue Lhomond, 75231 Paris Cedex 05, France. \\
$^2$ Universit\"at des Saarlandes, 66123 Saarbr\"ucken, Germany}

\begin{abstract}
Color centers in diamond are versatile solid state atomic-like systems suitable for quantum technological applications. In particular, the negatively charged silicon vacancy center (SiV) can exhibit a narrow photoluminescence (PL) line and lifetime-limited linewidth in bulk diamonds at cryogenic temperature. We present a low-temperature study of chemical vapour deposition (CVD)-grown diamond nano-pyramids containing SiV centers. The PL spectra feature a bulk-like zero-phonon line with ensembles of SiV centers, with a linewidth below 10 GHz which demonstrates very low crystal strain for such a nano-object. \end{abstract}

\maketitle

Photoluminescent impurities or impurity complexes in diamond, often termed color centers, are attractive solid-state, atom-like systems. Their stable, bright photoluminescence (PL), even for single centers, makes them ideal, e.g.\ as single photon sources or as fluorescent labels for bio-imaging. Their electronic spins form versatile sensors, e.g.\ for magnetic fields, electric fields, temperature and strain \cite{Bernardi2017}. Some defects stand out due to a highly-favorable combination of properties, among them the silicon vacancy (SiV) center. Negatively charged SiV centers (simply termed SiV center here) feature a very narrow, purely-electronic PL transition (zero-phonon-line, ZPL) at near infrared wavelengths ($\approx\,$740 nm, width $<$1 nm at room temperature), in which most of their PL (up to 88\%, \cite{Neu2011}) concentrates. Thus, SiV centers are promising candidates for fluorescent bio-labeling in cells \cite{Merson2013a} and enable efficient coupling to photonic structures \cite{Riedrichmoeller2011}. Recently,  SiV centers demonstrated their capability for high precision temperature sensing \cite{Nguyen2017}. Here, in contrast to thermometry based on nitrogen vacancy centers, SiVs enable all optical measurements without the need for microwave radiation. Moreover, SiV centers are the only color centers that have been identified in molecular-sized nanodiamonds \cite{Vlasov2014}. In quantum information, SiV centers stand out owing to their level scheme that enables ultra-fast spin-manipulation using ultra-short laser pulses \cite{Becker2016} and close to lifetime limited emission \cite{Sipahigil2014}. Recently, SiV centers provided potential qubit states with record-high coherence times exceeding 10 ms at temperatures below 500 mK \cite{Sukachev2017}. 

\begin{figure}[ht!!]
\centerline{\scalebox{0.18}{\includegraphics{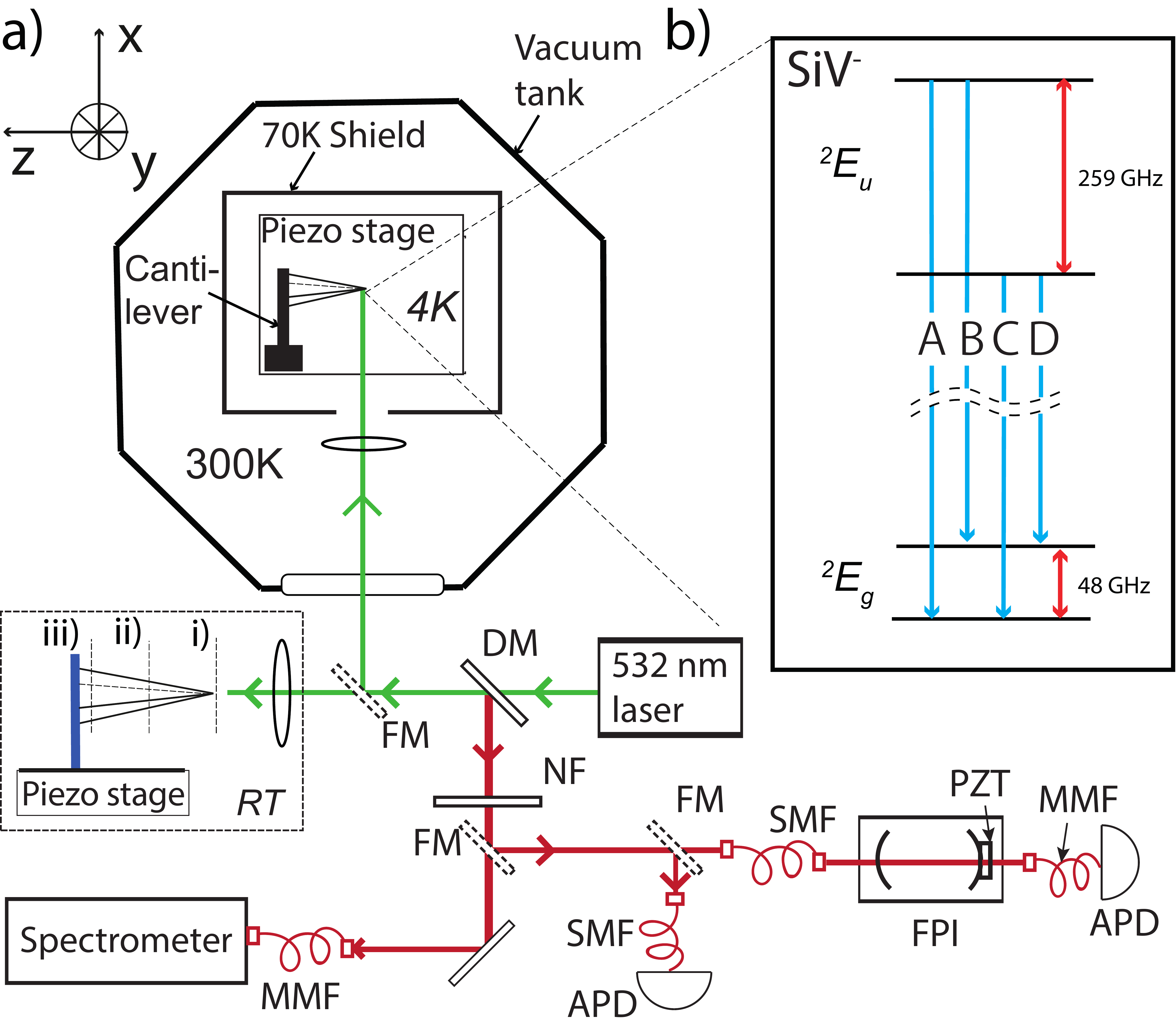}}}
\caption{a) Schematics of the experiment performed at 4 K and at room temperature (RT).    
DM=dichoic mirror, MMF/SMF=multi/single mode fiber, FM=flipping mirror, NA=numerical aperture, APD= avalanche photodiode, FPI=Fabry-P\'erot interferometer, PZT=piezo-electric transducer.
b) Electronic level scheme of the SiV$^-$ consisting of orbitally split ground and excited state doublets.}\label{setup}
\end{figure}

For many applications in sensing, labeling and quantum information, it is mandatory to incorporate color centers into nanostructures and nanodiamonds (NDs) to efficiently collect their PL, incorporate them into cells or use them for scanning probe-based sensing. Until now, incorporating SiV centers into diamond nanostructures remains partly challenging: First experiments on individual, bright SiV centers used NDs grown by chemical vapor deposition (CVD) on iridium substrates \cite{Neu2011}. Despite showing narrow ZPLs, individual SiVs in these NDs display ZPL wavelengths scattered within $\approx$ 10 nm together with significant spectral diffusion often severely limiting the ZPL line-width even for single centers at cryogenic temperatures \cite{Neu2013}. The same holds for SiVs in NDs obtained by milling CVD diamond films \cite{Neu2011a}. Only recently, CVD-grown NDs treated using oxygen plasma etching showed narrow ensemble (0.08 nm, 10 K) ZPLs revealing the four-line fine-structure in the SiV's level scheme \cite{Grudinkin2016}. Similar results, with resolvable fine structure (inhomogeneous broadening $<$ 250 GHz) were obtained using NDs grown via the high pressure high-temperature method and subsequently treated in hydrogen plasma \cite{Rogers2018}. In contrast, SiV centers in nanostructures obtained via plasma etching \cite{Evans2016, Marseglia2018} from single crystal diamond can show much narrower, almost transform limit lines \cite{Evans2016}. In single crystal diamond, the ZPL wavelength of SiV centers proved to be so stable, that photons from two separate, individual color centers interfere without any resonance tuning \cite{Sipahigil2014}, an observation not possible with any other solid-state emitter system to date. Using such structures, quantum optical switching on the single photon level was realized \cite{Sipahigil2016b}. The superior spectral properties of the SiV center in single crystals are attributed to its inversion symmetry, which renders it less prone to charge fluctuations (spectral diffusion \cite{Rogers2014b}) and leads to a narrow distribution of ZPL wavelengths.  

In this work, we obtain narrow ZPL from dense SiV center ensembles in nanostructures, namely diamond pyramids (DPs, from Artech Carbon). The DPs are grown using a dedicated CVD process in a bottom-up approach without the need for post-growth plasma treatment. The process straightforwardly creates many structures in one growth run. The DPs have favorable photonic properties and are routinely used to realize commercially available all-diamond atomic force microscope tips.      

\begin{figure}[ht!!]
\centerline{\scalebox{0.32}{\includegraphics{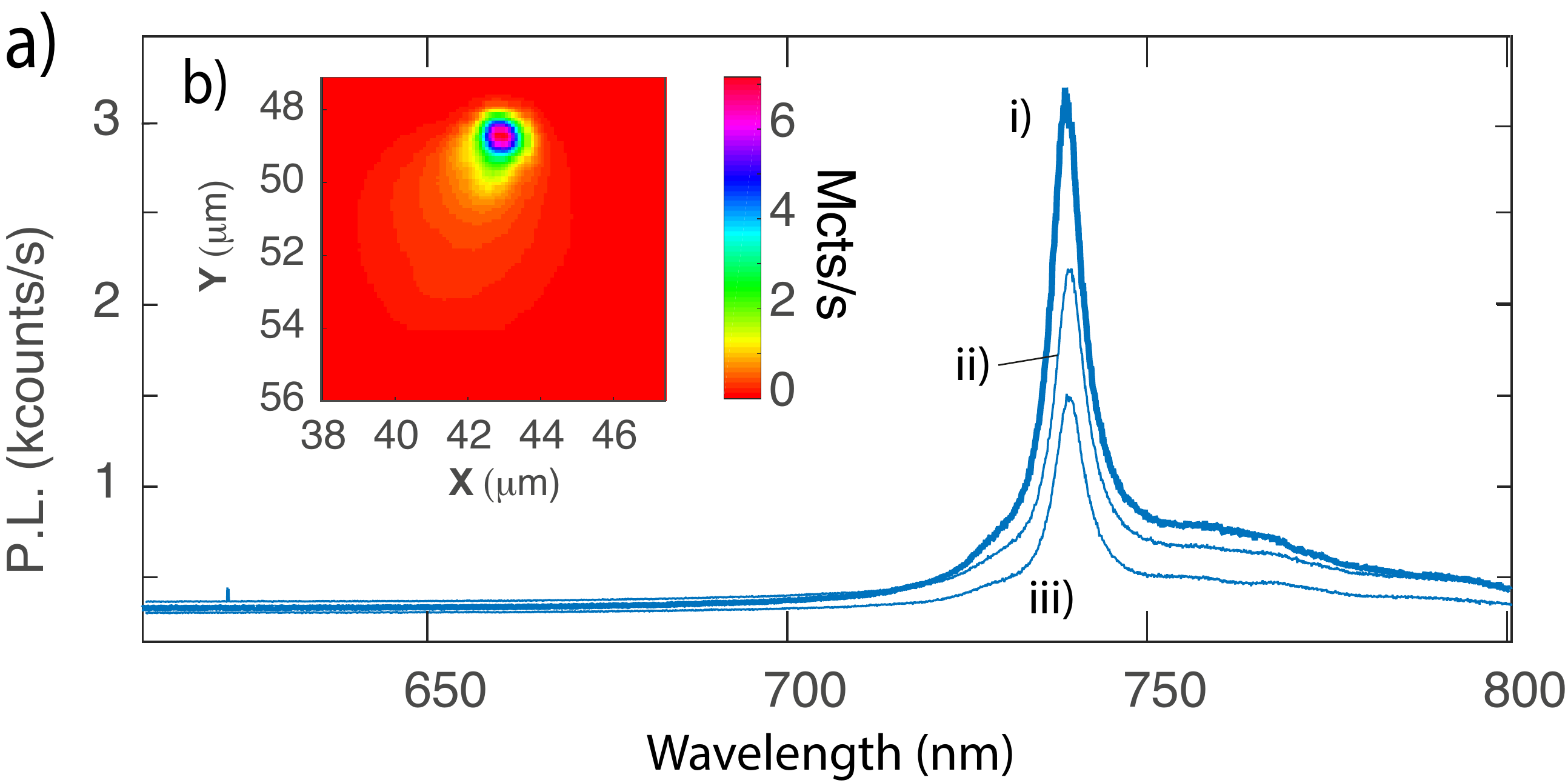}}}
\caption{a) Spectra taken at room temperature for different depth $z$ in the diamond pyramid. Trace i), ii) and iii) correspond to depths $z$=0, 4 and 10 $\mu$m  respectively. b) Confocal map taken at the apex of the pyramid in the wavelength range 600-800 nm.}\label{spectre}
\end{figure}

We use two types of DPs : a large collection of DP deposited on a silicon substrate and commercially available individual DPs that are fixed on a cantilever. 
The former (referenced D300) are manufactured mainly for their use as AFM probes.
All DPs have a square side of about 5 $\mu$m at the base and a height of 15 $\mu$m. The radius of curvature of the apex is less than 10 nm. 
  
We first study the photoluminescence properties of the DPs attached to the cantilever at room temperature using the home built confocal microscope depicted in Fig.~\ref{setup}-a). 
The DP main axis is positioned along the optical axis $z$ for these room temperature measurements. 
The microscope comprises an Olympus LCPlanFLN 50x objective with a numerical aperture (N.A) of 0.70. The DPs are mounted on a three-axis piezo-electric stage (PI NanoCube P-611.3). The excitation was performed with a 532 nm continuous diode laser (CPS532 Thorlabs) delivering a power of 100 $\mu$W. A dichroic mirror (DMLP550 Thorlabs) separates the red PL from the excitation laser and a notch filter centered at 532 nm removes the remaining excitation laser. The PL is coupled to a single fiber and detected by an avalanche photodiode (SPCM-AQRH-15 Excelitas Technologies). 

We have performed PL raster scans on three different DPs. An X-Y scan performed at the apex is shown in Fig.\ref{spectre}-b).
We then recorded the PL spectrum with a 1200 l/mm grating and an Andor camera (DU401A-BVF) for three depths $z$, as shown Fig.\ref{spectre}-a). Trace i) ii) and iii), correspond to $z=0$ (the apex of the DP), b) and c) to a distance $z=4$ and $z=10 \mu$m respectively, as sketched in figure 1-a). Here, to record the spectra, the PL is coupled into a multimode fiber and sent to the spectrometer.
At each depth $z$, a peak at around 740 nm with a width of around 5 nm appears, which corresponds to the ZPL of SiV centers. Signal at longer wavelengths corresponds to radiative emission assisted by phonons. 
These measurements demonstrate that SiV centers are mainly located at the apex of the DP, as also observed in \cite{Nelz2016}. 
Indeed the total PL drops from $7\times10^{6}$ counts/s to $5 \times 10^{5}$ counts/s when changing the depth from z=0 $\mu$m to z=15 $\mu$m.  When comparing with the results of \cite{Nelz2016}, taking into account the different numerical apertures, the SiVs in the DPs used here yield count rates on the same order of magnitude. 
We can thus use the estimation performed in \cite{Nelz2016}, and conclude that the SiV density lies in the tens of ppm range at the apex. 

Notably, in this batch of DPs, no PL lines due to other color centers are observed. 
In contrast to previous work  \cite{Nelz2016}, we here use DPs that have been grown in a nominally nitrogen-free process and thus obtain DPs with no nitrogen vacancy (NV) center based PL.
These measurements were performed on 4 different DPs and in fact only one DP spectrum indicated the presence of NV centers, although with a count rate that was an order of magnitude lower compared to the SiV.


\begin{figure}[ht!!]
\centerline{\scalebox{0.25}{\includegraphics{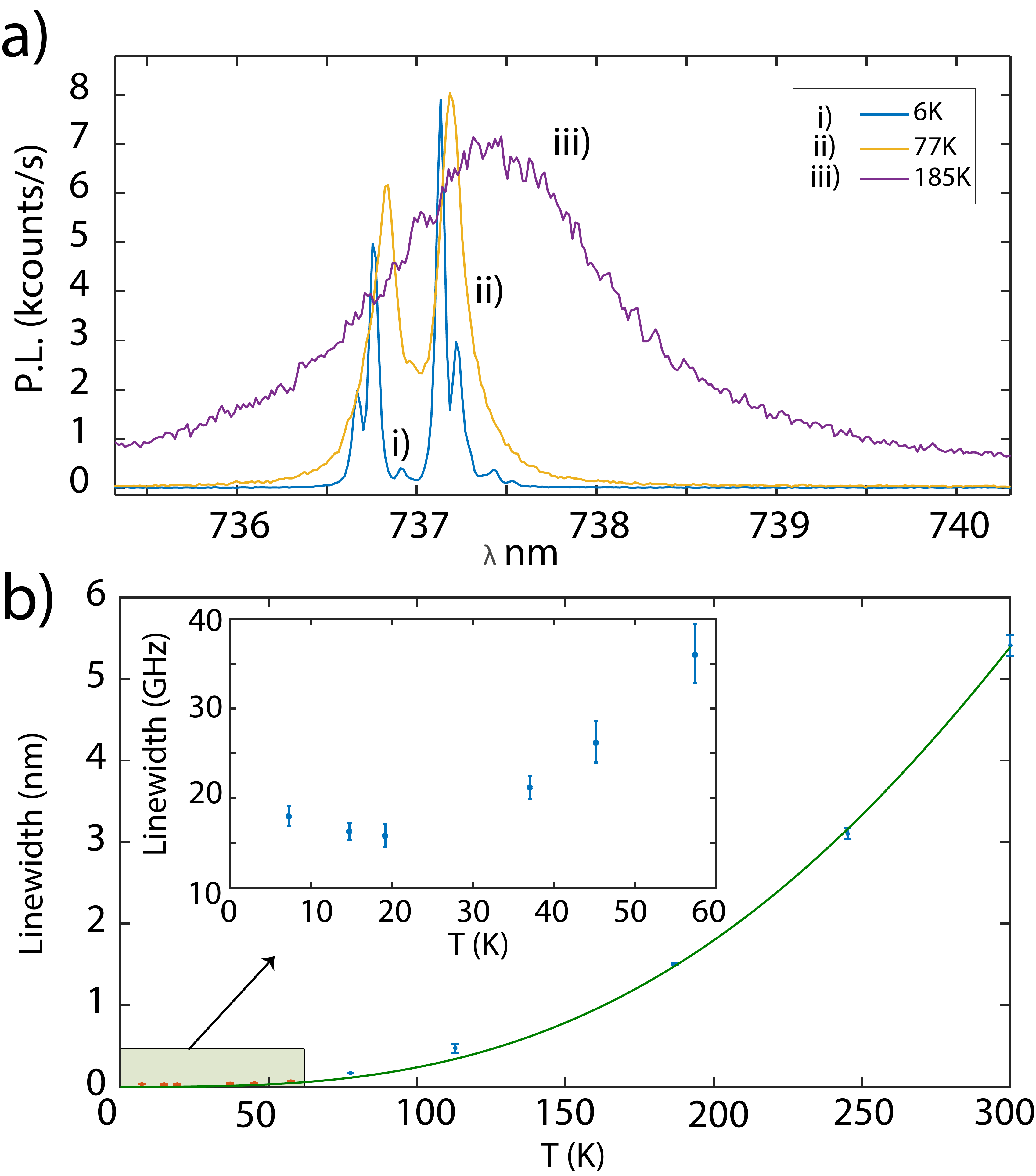}}}
\caption{a) SiV photoluminescence spectra for three temperatures : trace i) 6 K, ii) 77 K, and iii) 185 K. b) SiV zero-phonon linewidth as a function of temperature, measured using the spectrometer. The inset show low temperature measurements taken with the Fabry-P\'erot interferometer.}\label{Temp}
\end{figure}

We now proceed with low temperature measurements. The samples are placed in the cryostat shown Fig. \ref{setup}-a). This cryostat features two platforms: one platform at room temperature which holds the optical components and another one at 4K which holds the cold finger. These two parts are separated by a thermal screen and a quartz window and are both placed under vacuum. In principle, this system allows to bring the objectives very close to the sample in a cost effective way, since room temperature positioning stages and optics are usually cheaper.

The DP main axis is here positioned perpendicularly to the optical axis.
We use a RMS10X - 10X Olympus Plan Achromat objective, with a N.A. of 0.25. The sample is placed on a three axis stage (Attocube ANPxyz51).
We first measured the change in the emission spectrum with temperature using a 1800 l/mm grating. 
Fig. \ref{Temp}-a) i), ii) and iii) show three spectra, taken at temperatures of 6 K, 77 K and 185 K respectively, measured at the base of sample holder.
At temperatures below 60 K, four optical transitions in the ZPL of the SiV centers can be seen.
These main lines correspond to the four transitions of the SiV$^-$ centers of the predominant silicon isotope: $^{28}$Si (see Fig.~1-b)). Less pronounced peaks are also visible. Their positions and relative heights are consistent with the ZPL of the isotope of silicon: $^{29}$Si \cite{Dietrich2014}.
These spectra also feature the typical large blue shift of the ZPL transitions when the temperature is lowered due to the contraction of the crystal. A central wavelength of 737.95 nm is measured at room temperature while it is at 736.95 nm at 6 K.


\begin{figure}[ht!!]
\centerline{\scalebox{0.26}{\includegraphics{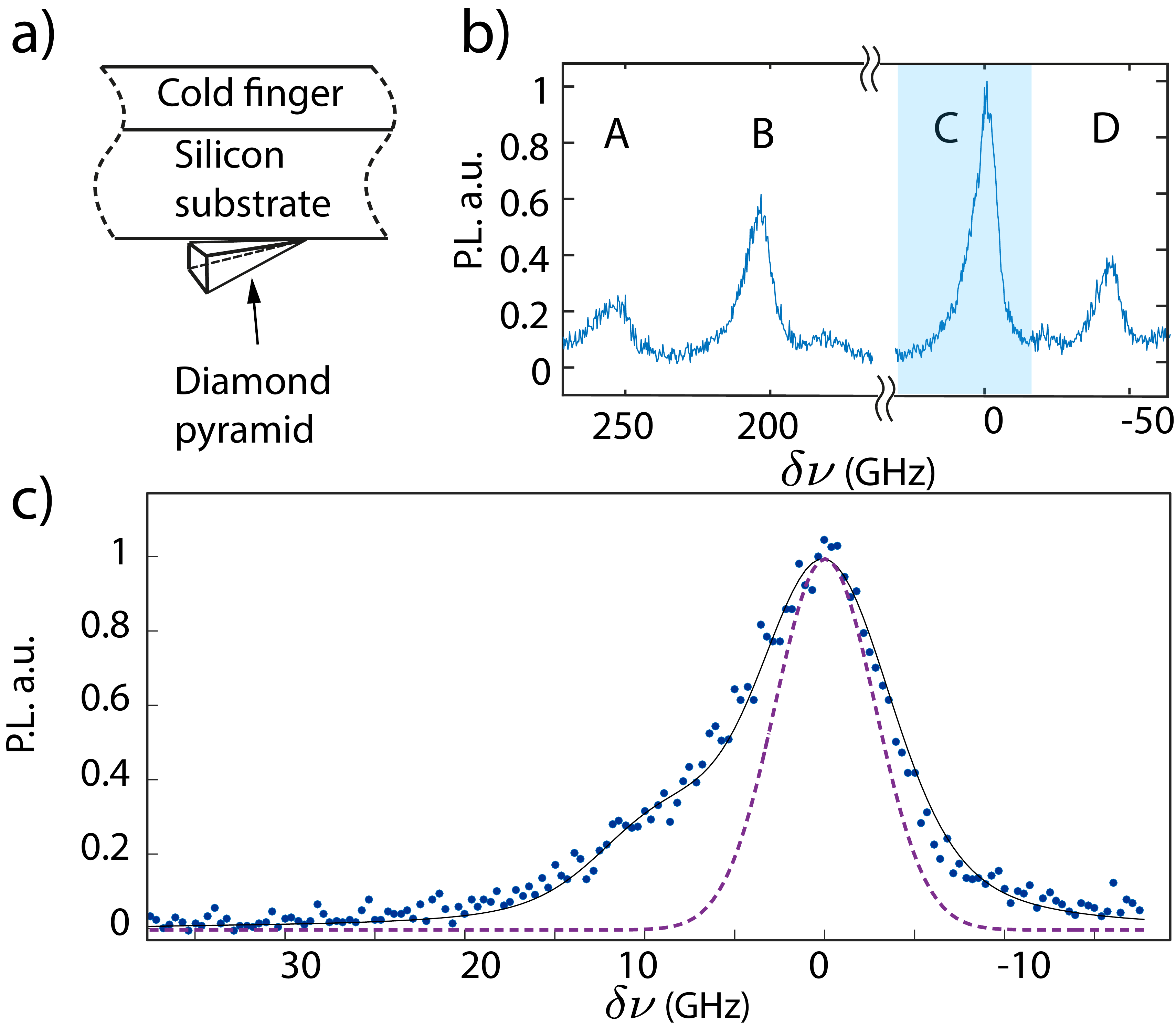}}}
\caption{a) The employed sample. b) Low temperature photoluminescence spectrum from a diamond pyramid deposited on a silicon substrate, measured using the Fabry-P\'erot interferometer. c)  Photoluminescence spectrum zoomed on the transition C of the SiV$^-$. The solid line is a fit to the data and the dashed curve the narrowest SiV ensemble ZPL we can extract from the fit (explanations in the text).
}\label{temp2}
\end{figure}

In Fig. \ref{Temp}-b), the linewidth $\Delta \lambda$ of the ZPL is plotted as a function of the temperature. Here, a $T^3$ scaling law (with a coefficient $a=2.447.10^{-7}$) fits well the data. In this range of temperature, the higher order coefficients in the polynomial expansion are negligible (the $T^5$ component is $b=-4.9.10^{-13}$).
The dependence of the linewidth with temperature depends on the impurity level in the sample and on the level of inhomogeneous broadening \cite{Neu2013}. 
Several scaling laws can thus be found in the literature, with generally a combination of $T^3$, $T^5$ or $T^7$ dependence.
The $T^3$ scaling observed here means that we can attribute the broadening dependence to fluctuating fields which are created as phonon modes 
change the distance between the color center and other defects in the crystal \cite{Hizhnyakov}.

The resolution of the spectrometer (0.05 nm) does not allow to measure the linewidth of each peak below 50 K however.
To go beyond the spectrometer resolution and measure the ZPL linewidths at low temperature, a scanning Fabry-P\'erot interferometer (FPI) was built. Its free spectral range (FSR) was chosen to be on the order of the distance between the two extremal emission lines. We measured it to be about 300 GHz. As depicted in Fig.~1-a), two concave mirrors (Layertec) with radii of curvature 75 mm and 150 mm and a reflectivity of 97\% are used, giving a finesse of 76 corresponding to a linewidth of 3.8 GHz. One mirror is mounted on a Jena Piezosystem actuator. 

The inset of Fig. 3-b) shows measurement of the linewidths obtained by fitting a Lorentzian profile to the obtained Bell shaped curves after the FPI. 
Gaussian fits indeed yield poor agreement with the data, letting us conclude that homogeneous broadening is dominant. 
A homogeneous linewidth varying of 17 to 20 GHz is measured in the temperature range 7-35 K. This is much larger than the homogeneous width given by the lifetime limit (around 200 MHz), indicating that the temperature measured at the DP apex is not below 10 K.

A possibility is that the cryostat was not able to cool the DPs below 20 K due to limited thermal contact between the DP base (with cross-section $5~\mu$m$^2$) and the cantilever, and between the cantilever and the cold finger. 
The DP is indeed glued at the the end of the cantilever, which has a small cross-section of $4\times 35~\mu$m$^2$, so a poor thermal conductivity. We thus chose to investigated DPs that were cast on a silicon substrate. As depicted in figure~\ref{temp2}-a), we could isolate single DPs. The DPs came from a former batch than the one , so they contain a significant amount of NV centers. We found that the SiV ZPLs are still very narrow at low temperatures and displayed similar spectra above 10 K. 
We then direct the SiV PL to the FPI to estimate linewidths below the spectrometer resolution. The results of the measurements are shown Fig. \ref{temp2}-b).
In contrast to the FPI spectra obtained with the previous DPs, the four lines are almost a factor of two narrower, yielding ZPL full widths at half-maximum ranging from 8 to 9 GHz, after deconvolution with the FPI response. 

As a comparison, this value is on a par with the 9 GHz width observed in \cite{Neu2013} on ensemble of SiV centers in bulk homo-epitaxial films.  
It is also smaller than the narrowest reported linewidth measured with ensembles of SiV in nanodiamonds \cite{Grudinkin2016}. 
There, the growing was realized using nanodiamond seeds with further CVD growth and reactive ion etching (RIE).  
Another approach is to implant SiVs using focused ion beam in a low strain diamond substrate followed by several annealing steps.
In \cite{Evans2016}, the ZPL width of SiV centers was measured on 13 centers. The resulting ensemble width, estimated using half of them, was 15 GHz.   
Here, observing less than 10 GHz linewidths on ensembles of SiV centers underlines the very small strain in the DP which is usually very challenging to reach and requires specific growing and etching conditions.

As opposed to our previous measurements on DPs glued to cantilevers, all the ZPL peaks now display an asymmetric profile, with a tail dragging to low wavelengths. This could be explained as a varying strain together with a SiV density along the DP main axis. 
The fact that the profile on all transitions is the same points towards axial strain on all SiV centers. 
A FPI scan around the C transition is shown in Fig. \ref{temp2}-c). The simplest employed fit we used, a double Gaussian convoluted by the FPI Lorentzian response, reproduces well the data, which implies negligible homogeneous broadening.
The frequencies around which the count rate is large is likely to come from SiV centers at the apex. 
The sharp decrease of the ZPL at high wavelengths may thus be used to estimate the spectral width of SiVs that reside at the DP apex.
The corresponding Gaussian curve used in the full fit of Fig. \ref{temp2}-c) can be extracted. It is plotted in Fig. \ref{temp2}-c) (dashed line) and features a full-width at half maximum of 6 GHz, which may result from spectral diffusion or local strain. Further studies will be conducted to verify this conjecture. 

In conclusion, we presented low temperature studies of CVD-grown diamond nano-pyramids containing SiV centers. The pyramids are grown in a bottom-up approach without the need for plasma treatment, feature narrow ($8-10$ GHz) characteristic bulk-like structures with ensembles of SiV centers together with a negligible NV centers density, making them attractive for quantum optics and sensing experiments. \\

\acknowledgments
We would like to thank Vitali Podgurski and Julien Paris from MyCryoFirm for technical support
GH acknowledges funding by the French National Research Agency (ANR) through the project SMEQUI.
EN acknowledges funding via a NanoMatFutur grant of the German Ministry of Education and Research (FKZ13N13547) as well as a PostDoc Fellowship by the Daimler and Benz Foundation

\end{document}